\documentclass[letterpaper, 10 pt, conference]{ieeeconf}  

\IEEEoverridecommandlockouts
\usepackage{amsfonts}
\usepackage{graphicx}
\usepackage{cite}
\usepackage{color}
\input{steve.lib}

\newcommand{\pf}{\noindent{\bf Proof }}

\title{A Hybrid Observer  for a  Distributed Linear System with a Changing Neighbor Graph}

\author{L. Wang$^{1}$,   A. S. Morse$^{1}$, D. Fullmer$^{1}$,
and J. Liu$^{2}$
\thanks{This work was  supported by  National Science Foundation grant n. 1607101.00 and  US Air Force grant n. FA9550-16-1-0290.}
\thanks{$^{1}$ L. Wang, A. S. Morse and D. Fullmer are with the Department
of Electrical Engineering, Yale University, New Haven, CT, USA.
        {\tt\small \{lili.wang,  as.morse, daniel.fullmer\}@yale.edu }}%
\thanks{$^{2}$ J. Liu  is with Coordinated Science Laboratory, University
 of Illinois at Urbana-Champaign,  Champaign, IL, USA.
        {\tt\small \{jiliu\}@illinois.edu }}%
        }
\begin{document}

 \maketitle
 \thispagestyle{empty}

\begin{abstract} A hybrid  observer is described for  estimating the state
of an $m>0$ channel, $n$-dimensional, continuous-time, distributed
linear system of the form $\dot{x} = Ax,\;y_i =
C_ix,\;i\in\{1,2,\ldots, m\}$. The system's state $x$ is
simultaneously estimated  by $m$ agents assuming each agent $i$
senses  $y_i$ and receives appropriately defined data from each of
its current neighbors. Neighbor relations are characterized by a
time-varying directed  graph $\mathbb{N}(t)$ whose
 vertices correspond to agents and whose arcs depict neighbor relations.
Agent $i$  updates its   estimate  $x_i$  of $x$ at ``event times''
$t_1,t_2,\ldots $ using a local observer and a  local parameter
estimator. The  local observer  is a continuous time  linear system
whose
 input is $y_i$
and whose output $w_i$ is an asymptotically correct estimate of
$L_ix$ where $L_i$ a matrix with kernel equaling the unobservable
space of $(C_i,A)$. The local  parameter estimator is a recursive
algorithm  designed to estimate, prior to each event time $t_j$,
 a constant parameter $p_j$ which satisfies the linear equations
$w_k(t_j-\tau) = L_kp_j+\mu_k(t_j-\tau),\;k\in\{1,2,\ldots,m\}$,  where $\tau$ is a small positive constant and  $\mu_k$ is the state estimation error of local
observer $k$.
Agent $i$  accomplishes this   by iterating its parameter estimator state $z_i$,  $q$ times within the interval $[t_j-\tau, t_j)$,   and by making use of the state of  each of its neighbors' parameter estimators
     at each iteration.  The updated
  value of $x_i$ at event time $t_j$ is then $x_i(t_j) = e^{A\tau}z_i(q)$.
Subject to  the assumptions that
   (i) none of the $C_i$ are zero,
(ii) the   neighbor graph $\mathbb{N}(t)$ is strongly connected for all time, (iii)
the system whose state is to be estimated is jointly observable, (iv) $q$ is sufficiently large
 and nothing more, it is shown that each estimate $x_i$ converges to $x$ exponentially fast as $t\rightarrow \infty$
  at a rate which can be controlled.

\end{abstract}

\section{Introduction}

In a recent paper \cite{Lilimorse2017}, a distributed   observer was
described for  estimating the state of an $m>0$ channel,
$n$-dimensional,   continuous-time, jointly observable linear system
of the form $\dot{x} = Ax,\;y_i = C_ix,\;i\in \{1,2,\ldots, m\}$.
The state $x$ is   simultaneously estimated by $m$ agents assuming
each agent $i$ senses  $y_i$ and receives the state  of each of its
neighbors' estimates. An attractive feature of the  observer
described in \cite{Lilimorse2017} is that it is able to generate an
asymptotically correct estimate of $x$ at a pre-assigned exponential
rate,
 if each agent's neighbors do not change with time and the neighbor graph characterizing neighbor relations is strongly connected.
However,  a   shortcoming of this observer is that it
is unable to function correctly if the
 network changes with time. Changing neighbor graphs will typically occur if the agents are mobile.
 A second shortcoming of the
 observer described in \cite{Lilimorse2017}  is that it is ``fragile.''
 By this we mean that the observer is not
able to cope
 with the situation
when an agent's neighbors change.  For example,
 if because of a component failure, a loss of battery power or some other reason, an agent drops out of the network,
what remains of the observer will typically  not be able to perform correctly and   may  become unstable, even if
what is left is still  a jointly observable  system  with a strongly connected neighbor graph.
The aim of this paper is to describe a new
type of  observer which overcomes these difficulties.

\section{The Problem}\label{prob}
We are interested in a   network of $m>0$ autonomous agents
labeled $1,2,\ldots, m$  which are able to receive information from
their neighbors where by the  {\em neighbor} of agent $i$ is meant
any other agent in agent $i$'s reception range.
 We write
 $\scr{N}_i(t)$ for the set of labels of agent $i$'s neighbors  at real time $t\in[0,\infty)$ and we take agent $i$ to be a neighbor of itself.
  Neighbor relations at time $t$  are
 characterized  by a directed graph $\mathbb{N}(t)$ with $m$ vertices and a set of arcs defined so that there is an arc from
  vertex $j$ to vertex $i$  whenever agent $j$ is a   neighbor of agent $i$. Each agent $i$
  can sense a continuous-time signal $y_i\in\R^{s_i},\;i\in\mathbf{m}=\{1,2,\ldots, m\}$, where \begin{eqnarray}y_i &=&C_ix,\;\;\;i\in
  \mathbf{m}\label{sys1}\\\dot{x} &= &Ax\label{sys2}
\end{eqnarray}
and $x\in\R^n$.  We assume throughout that $C_i \neq 0,\;i\in\mathbf{m}$, and that the system defined by \rep{sys1}, \rep{sys2} is {\em jointly observable};
     i.e.,  with
    $C = \matt{C_1' &C_2' &\cdots &C_m'}'$, the matrix pair $(C,A)$ is observable.
The problem of interest is to develop ``private estimators'',
one for each agent, which enable each agent to obtain an asymptotically correct estimate of $x$.

\subsection{Background} Distributed state estimation problems  have been under study in one form or another for years.
In many cases
system and measurement noise are components of the problem considered and some form of Kalman filtering is proposed.
The literature on this subject is vast, and many specialized results exist; see for example,
\cite{martins,KhanAli2011ACC,Kim2016CDC,MitraPurdue2016,shamma,carli,xxx,saber2,bullo.observe} and the many
references cited therein. However, to the best of our knowledge, the specific problem we have posed has not been solved without
imposing restrictive assumptions.
One reason for this we think is because most approaches rely on estimators which are time-invariant linear systems.
 We believe that the problem
posed, involving a time-varying neighbor graph, cannot be solved without qualification, with a time invariant linear system. It would be
especially useful to know whether or not this  conjecture is  true.

\section{Observer}\label{ob}

The idea we  are about to present makes use of to two familiar types of systems.  The first type is a  classical observer;
such systems
  enable each agent to generate
 an asymptotically  correct estimate of the ``part of $x$ which is  observable'' to that particular agent.
  The second type of system is  a parameter estimator; using parameter estimators
 enables each agent to generate an asymptotically correct estimate of $x$ frozen at a specified  time instant
 by viewing $x$ at that time as a fixed parameter.
    The judicious  combination of these two types of systems provides a straightforward, easy to analyze solution
 to the problem we have posed and it is surprising that the idea has not been suggested before.

The observer to be described consists of $m$  estimators, one for
each agent. Agent $i$ generates an estimate $x_i$ of  $x$ with its
private estimator  $\mathbb{E}_i$ which is  a hybrid dynamical
system consisting of
 a ``local observer'' and a  ``local parameter estimator.'' Each $x_i$ is updated at {\em event times} $t_1,t_2,\ldots $
  where $t_j = jT,\;j\geq 1$, and $T$ is a pre-selected positive real number.
 Between event times, each $x_i$ satisfies $\dot{x}_i=Ax_i$. Agent $i$'s {\em local observer} is  a continuous time  linear system  whose
 input is $y_i$
and whose output $w_i$ is an asymptotically correct estimate of
$L_ix$ where $L_i$ a matrix with kernel equaling the unobservable
space of $(C_i,A)$. The  computations needed to update  each agent's
estimate of $x$ at event time $t_j$ are  carried out
   over the time interval $[t_j-\tau, t_j)$; here
$\tau$ is a positive number smaller than $T$ which is chosen large enough  so that the
  computations required to  update each agent's estimate  can be completed in $\tau$ time units. Agent $i$'s {\em local parameter estimator}
is a recursive algorithm  designed to estimate on each interval
   $[t_j-\tau, t_j)$, a constant parameter $p_j$ which satisfies the linear equations
$w_k(t_j-\tau) = L_kp_j+\mu_k(t_j-\tau),\;k\in\mathbf{m}$,  where $\mu_k$ is the state estimation error of local
observer $k$.
Agent $i$  accomplishes this   by iterating its parameter estimator state $z_i$,  $q$ times within the interval $[t_j-\tau, t_j)$,   and by making use of the state of  each of its neighbors' parameter estimators
     at each iteration.  The updated
  value of $x_i$ at event time $t_j$ is then $x_i(t_j) = e^{A\tau}z_i(q)$.

\subsection{Estimator $\mathbb{E}_i$}

In this section we give a more detailed description of agent $i$'s private estimator. As just stated,
the estimator consists of a local observer and a local parameter estimator.

\subsubsection{Local Observer $i$}
Recall that the unobservable space of $(C_i,A)$, written $[C_i|A]$,
is the largest $A$-invariant subspace contained in the kernel of
$C_i$. Set $n_i = n - \dim([C_i|A])$ and let $L_i$ be any $n_i\times
n$ matrix  whose kernel is $[C_i|A]$. Then as is well known, the
equations $C_i = \bar{C}_iL_i$ and $L_iA = \bar{A}_iL_i$ have unique
solutions $\bar{C}_i$ and $\bar{A}_i$ respectively and
$(\bar{C}_i,\bar{A}_i)$ is an observable matrix pair. By a {\em
local observer} for agent $i$ is meant any $n_i$ dimensional system
of the form \eq{ \dot{w}_i = (\bar{A}_i+
K_i\bar{C}_i)w_i-K_iy_i\label{po}} where $K_i$ is a matrix to be
chosen. It is easy to verify that the local observer estimation
error $\mu_i \dfb w_i-L_ix$ satisfies
$$\mu_i(t) = e^{(\bar{A}_i+ K_i\bar{C}_i)t}(w_i(0)-L_ix(0)),\;\;\;t\in[0,\infty)$$
Moreover, since $(\bar{C}_i,\bar{A}_i)$ is observable, $K_i$ can be
selected so that $\mu_i(t) $ converges to $0$ exponentially fast at
any pre-assigned rate. We assume that each $K_i$ is so chosen. Since
\eq{w_i(t) = L_ix(t) +\mu_i(t),\;\;\;t\in[0,\infty)\label{e1}} $w_i$
can be thought of as an asymptotically correct estimate of $L_ix$.

\subsubsection{Local Parameter Estimator $i$}

The starting point for the development of the local parameter estimators is the observation that for each event time $t_j$,
the system of equations
 \eq{w_i(t_j -\tau) = L_ip_j +\mu_i(t_j-\tau),\;\;i\in\mathbf{m}\label{noise}}
  has a unique solution, namely $p_j= x(t_j-\tau)$. This is a consequence of
 \rep{e1} and the joint observability assumption. It is useful to think of the estimation of $p_j$ as a parameter
  estimation problem. One
 algorithm for computing $p_j$ which would give an asymptotically correct result in an infinite number of steps
  if each  $\mu_k(t_j-\tau)$ were zero, is the algorithm described in \cite{liliacc16}.
   In this paper we will make use of this algorithm   but will
  only iterate $q>0$ steps where $q$ is an integer-valued
 design constant which is  chosen large enough to ensure exponential convergence; we assume that
 the local processers are sufficiently fast so that
 each can execute
  $q$ iterations in
 $\tau$ time units. The {\em local parameter estimator} for agent $i$   as  defined as follows.
 For each event time $t_j$,
\begin{eqnarray} && \hspace{-1cm} z_i(0) = x_i(t_j-\tau)\label{l1}
\\ \nonumber &&  \hspace{-1cm}  z_i(k)= \bar{z}_i(k-1)\\
& & -Q_i(L_i\bar{z}_i(k-1) - w_i(t_j-\tau)),\;\;k\in\mathbf{q}
\label{l2}
\\ &&  \hspace{-1cm}  x_i(t_j)=e^{A\tau}z_i(q)\label{l3}\end{eqnarray}
where $ k \in \mathbf{q}\dfb\{1,2,\ldots,q\}$, $Q_i = L_i'(L_iL_i')^{-1}$, and
$$\bar{z}_i(k-1) = \sum_{ {s\in\scr{N}_i(\tau_k)}}z_s(k-1).$$
Here
 $$\tau_k
=t_j-\left (1-\frac{(k-1)}{q}\right )\tau $$ and
 $m_i(k)$
is the number of labels in $\scr{N}_i(\tau_k)$.
Note that the same symbols $z_i(k-1)$ and $\tau_k$ are used on each interval $[t_j-\tau,t_j)$, $j\geq 1$, without
 explicitly showing their dependence on $j$.

One way to modify the above algorithm without changing its essential features, is to redefine each $Q_i$ in \rep{l2}
as $Q_i = L_i'G_i$ where $G_i$ is any positive definite matrix for which the spectrum of
 $L_i'G_iL_i$ is contained in the open half interval $(-1,1]$. It is known that with this modification,
 the algorithm has the same convergence properties as the original but perhaps with a faster convergence rate if
 the $G_i$ are chosen appropriately \cite{liliacc16}.




\section{Main Result}\label{MR}

 The main result of this paper is as follows.

\begin{theorem} \label{thm:1} Suppose that \rep{sys1}, \rep{sys2} is jointly observable, that $C_i\neq 0,\;i\in\mathbf{m}$,
 and that the neighbor graph
$\mathbb{N}(t)$ is strongly connected for all $t\in[0,\infty)$. Then
for appropriately chosen $T,\tau, q,$ and $K_i,\;\;i\in\mathbf{m}$,
there exist positive constants $g$ and $\lambda$ for which the
following statement is true. For each  initial process state $x(0)$,
each initial local observer state $w_i(0),\;i\in\mathbf{m}$, and
each estimate $x_i(0),\;i\in\mathbf{m}$, \eq{|x_i(t)-x(t)|_2\leq
e^{-\lambda t}(\delta_x +\delta_{\mu}),\;t\geq
0,\;\;\;i\in\mathbf{m}} where
$$ \delta_x =\max_{i\in\mathbf{m}}|x_i(0)-x(0)|_2,\; \delta_{\mu} =g\max_{i\in\mathbf{m}}|w_i(0)-L_ix(0)|_2$$
and $|\cdot|_2$ is the standard  two norm.
\label{main}\end{theorem}

It is possible to give a formula for $\lambda $.  Towards this end,
 assume that $q$ has been chosen large enough so that
\eq{q>\left (1+\frac{\zeta T}{\ln \left (\frac{1}{\gamma }\right )}\right )((m-1)^2 +1)\label{q}}
where $\zeta $ is the largest eigenvalue of $\frac{1}{2}(A+A')$, and $\gamma$ is the  positive number
defined by  \rep{gamma} in Proposition \ref{prop:contraction}; note that $\gamma  $ is less than $1$ and  depends only on $m$ and
the $L_i$ which in turn depend only on $A$ and the $C_i$.
Next let $\omega $ be any positive number such that \eq{\omega +\zeta > \frac{r}{T}\ln \left  (\frac{1}{\gamma}\right )\label{omega}}
where $r$ is the unique integer quotient of $q$ divided by $(m-1)^2+1$. Assume that the $K_i$
used in the definitions of the local observers, have been chosen so that each local observer  estimation error decreases
in norm as fast as $e^{-\omega t}$ does. A formula for $\lambda$ is then
\eq{ \lambda  = \frac{r}{T}\ln \left  (\frac{1}{\gamma}\right ) - \zeta\label{lambda}}
Note that so long as \rep{q} holds, $\lambda >0$.  Note also
 that  the formula for $\gamma $ in Proposition \ref{prop:contraction}
is conservative and consequently so is the above formula for $\lambda$.


\subsection{Example}\label{EX}

The following example illustrates how
the observer performs when applied to an unstable system.
Consider the three channel, four-dimensional, continuous-time system
described by the equations  $\dot{x}=Ax,\; y_i=C_ix,\; i\in \{1,\;2,\;3\}$, where
$$A=\matt{ 0& 0.4& 0 & 0\cr 0&0&0&0\cr 0 &0 & 0 & 2\cr
0 & 0 & -2 & 0.2} \hspace{.15in}\begin{array}{lll}C_1&=&\matt{1& 0 & 0 & 0}\\\\
C_2&=&\matt{0& 1 & 0& 0}\\\\
C_3&=&\matt{0& 0 & 1&1}\end{array}$$ Note that $A$ has two eigenvalues at $0$ and  a pair of complex eigenvalues at
$0.1\pm j2.00$.
 While the system is jointly observable, no single pair $(C_i,A)$  is observable.

For this example $\scr{N}_1 =\{1,2\}$, $\scr{N}_2 = \{1,2,3\}$, $\scr{N}_3 = \{2,3\}$,
 $T=1$, $\tau=0.5$, $\gamma = 0.975 $ and $\zeta = 0.2$. To satisfy \rep{q}, $q$ is chosen as $q=45$
 and $r=9$.
To satisfy \rep{omega}, $\omega $ is chosen as $\omega =2$. The local observers for the three agents are constructed
 using the  following matrices.

\noindent For agent 1: \begin{eqnarray*} && \bar{C}_1=\matt{0&1},\;\;\;
\bar{A}_1=\matt{0 & 0\cr 0.4 & 0},\\ && L_1 = \matt{0 & 1 & 0 & 0\cr
1 & 0 & 0 & 0 },\;\;\;K_1 = -\matt{20\cr 6}\end{eqnarray*}
 For agent 2:
$$ \bar{C}_2=1,\;\;\;  \bar{A}_2=0,\;\;\; L_2 =\matt{0& 1 & 0 & 0},\;\;\;K_2 =-2$$
For agent 3: \begin{eqnarray*} &&\bar{C}_3=\matt{0&
\sqrt{2}},\;\;\;\bar{A}_3=\matt{0.1 & -1.9\cr 2.1 & 0.1},\\
&& L_3= \matt{0 & 0 & -\frac{\sqrt{2}}{2} & \frac{\sqrt{2}}{2}\cr 0
& 0 & \frac{\sqrt{2}}{2} &
\frac{\sqrt{2}}{2}},\;\;\;K_3=-\matt{0.85\cr 3.68}\end{eqnarray*}
In all three cases the convergence rate is $2$. Finally, for this
example \rep{lambda} gives an overall convergence rate of $\lambda
=0.025$.

\begin{figure}[h]
\centerline{\includegraphics [ height =4.in]{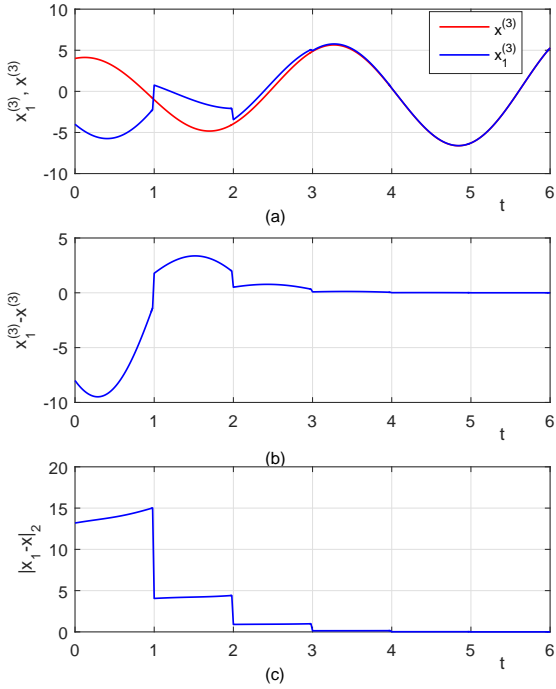}}
\caption{Simulation Results}
   \label{zop}
\end{figure}

This system was simulated with
 $x(0) = \matt{3 & 2 & 4 &1}' $ as the initial state of the process,  $w_1(0) = \matt{2 & 4}'$, $w_2(0) = \matt{3}$, and
 $w_3(0) = \matt{1 & 2}'$ as the initial states of the three local observers,
 and  $x_1(0) = x_2(0) = x_3(0) =  -\matt{4 & 4 & 4& 4}'$ as the initial estimates of the three local estimators.
The  two traces in Figure \ref{zop}a show the  simulation result for
the third components of $x_1 $ and $x$ respectively, namely
$x_1^{(3)}$ and $x^{(3)}$. The  trace in Figure \ref{zop}b shows the
error  $x_1^{(3)}-x^{(3)}$ while the trace  in Figure \ref{zop}c,
shows the two-norm of the error $x_1-x$.

\begin{figure}[h]
\centerline{\includegraphics [ height =4.in]{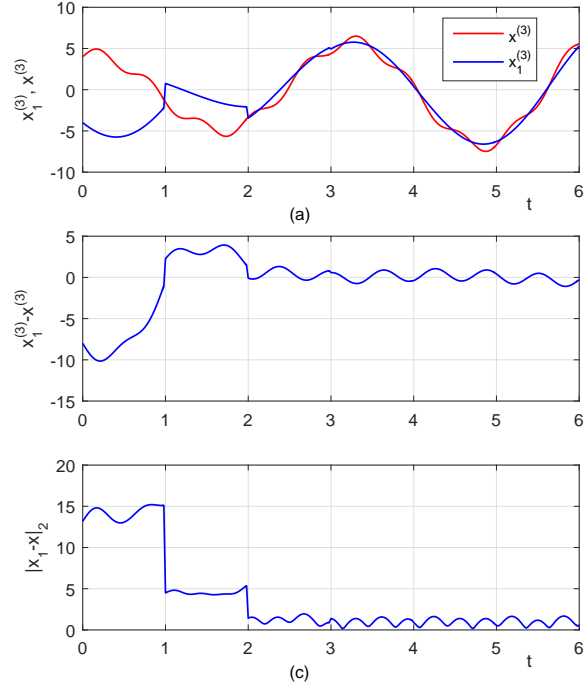}}
\caption{Simulation Results with System Noise}
   \label{zop2}
\end{figure}

To study the effect of an unmeasured disturbance driving  the process dynamics,
a second simulation was performed using   the same observer as above
 applied to the modified state equation system $\dot{x} = Ax +b\nu$  where $b = \matt{1 & 1 & 1&1}'$ and $\nu = 7\cos 10t.$
The resulting traces are shown in Figures \ref{zop2}a, \ref{zop2}b,
and \ref{zop2}c respectively.

\section{Analysis}\label{anal}

Fix $j>0$.  Our immediate aim is to analyze the behavior of the parameter estimators on the time interval $[t_j-\tau,t_j)$ .
Towards this end, for each $i\in\mathbf{m}$  let $\epsilon_i$ denote the parameter estimation error
$\epsilon_i(k) = z_i(k)-p_j,\;k\in\{0,1,\ldots,q\}$.
We claim that
\begin{eqnarray}  \hspace{-0.8cm}\epsilon_i(0)    &= &
e^{A(T-\tau)}(x_i(t\!_{j\!-\!1})-x(t\!_{j\!-\!1}))\label{pp1}\\
\nonumber \hspace{-0.8cm}
\epsilon_i(k ) &=& \frac{1}{m_i(k)}P_i\sum_{s\in\scr{N}_i(\tau_k)}\epsilon_s(k-1) \\
&\ &   +Q_i\mu_i(t_j-\tau), \hspace{0.5cm}k \in \mathbf{q}\label{pp2}\\
   \hspace{-0.8cm} x_i(t_j)-x(t_j) & =& e^{A\tau}\epsilon_i(q)\label{pp3}
\end{eqnarray}
where $t_0 = 0$, and $P_i$ is the orthogonal projection matrix $P_i =I- L_i'(L_iL_i')^{-1}L_i$. 
To establish \rep{pp1}, note that $\epsilon_i(0) = x_i(t_j-\tau)-x(t_j-\tau)$ because of \rep{l1},
the definition of $\epsilon_i$ and the fact that $p_j =x(t_j-\tau)$; \rep{pp1} follows at once. The recursion in \rep{pp2}
is an immediate consequence of \rep{noise} and  \rep{l2}. To establish \rep{pp3}, note that
$x_i(t_j)-x(t_j) = e^{A\tau}z_i(q)-x(t_j)$ because of \rep{l3}; but $x(t_j) = e^{A\tau}x(t_j-\tau) = e^{A\tau}p_j$.
Therefore \rep{pp3} is true.

To proceed define $\widehat{x} = \matt{x_1' & x_2' &\cdots & x_m'}'$,  $\bar{x} = \matt{x' & x' &\cdots &  x'}'$
and   $\epsilon = \matt{\epsilon_1' & \epsilon_2' &\cdots & \epsilon_m'}'$.
Write   $F_j(k)$ for the ``flocking matrix'' determined by $\mathbb{N}(\tau_k)$; i.e.,
$F_j(k) = D^{-1}_{\mathbb{N}(\tau_k)}A'_{\mathbb{N}(\tau_k)}$
where $D_{\mathbb{N}(\tau_k)}$ is the diagonal matrix of in-degrees
of the vertices of $\mathbb{N}(\tau_k)$ and $A_{\mathbb{N}(\tau_k)}$ is the adjacency matrix of $\mathbb{N}(\tau_k)$.
Then it is easy to verify that
\begin{eqnarray*}
\epsilon(0)  & =&e^{\bar{A}(T-\tau )}(\widehat{x}(t_{j-1}) - \bar{x}(t_{j-1}))\\
\epsilon(k)& = &P(F_j(k)\otimes I)\epsilon(k-1)
\\ &\ & +Q\mu(t_j-\tau),\;\;\;k\in\mathbf{q}\\
\widehat{x}(t_j)-\bar{x}(t_j)& =&
e^{\bar{A}\tau}\epsilon(q)\end{eqnarray*} where $\mu =\matt{\mu_1' &
\mu_2' &\cdots & \mu_m'}'$, $\bar{A} ={\rm
block\;diagonal}\{A,\;A,\;\ldots,\;A\}$, $P = {\rm
block\;diagonal}\{P_1,\;P_2,\;\ldots,\;P_m\}$,
 $Q = {\rm block\;diagonal}\{Q_1,\;Q_2,\;\ldots,\;Q_m\}$,
$\otimes$ is the
Kronecker product, and $I$ is the $n\times n$ identity matrix.
From these equations it follows that
\begin{equation}
\label{eq:errordynamic} \widehat{x}(t_j)-\bar{x}(t_j) =
\Omega_j(\widehat{x}(t_{j-1}) - \bar{x}(t_{j-1}))  + \Theta_j\mu(t_j
-\tau)
\end{equation}
where \begin{eqnarray}\nonumber \Omega_j &= &
e^{\bar{A}\tau}P(F_j(q)\otimes I) \\ &\
&P\!(\!F_j(q\!-\!1\!)\!\otimes \! I\!)\!\cdots \!
P\!(F_j(1)\!\otimes \!
I)e^{\bar{A}(T\!-\!\tau)}\label{big1}\end{eqnarray} and
\begin{eqnarray} \nonumber \hspace{-12mm}&&  \Theta_j =e^{\bar{A}\tau} \\ \hspace{-12mm}&&  \left(\! \sum_{s=2}^q \!P\!(\!F\!\!_j(q)\!\otimes \!I\!)\!
P\!(F\!\!_j(q\!-\!1\!)\!\otimes \!I\!)\!\cdots
\!P\!(\!F\!\!_j(s)\!\otimes \!I\!)\!+\!I\!\! \right)\! \!Q
\label{big2}\end{eqnarray} Suppose that  with some suitably defined
norm, the norm of each $\Omega_j$  is less than one and $\Theta_j$
is uniformly bounded as a function of $j$. Then the sequence
$\widehat{x}(t_j)-\bar{x}(t_j), j\geq 1$ will converge
 to zero at a exponential rate, because the sequence $\mu_j(t_j - \tau),\;j\geq 1$ converges to zero at an exponential rate.
 Because of this and  the fact that the time between successive event times is $T$, $\widehat{x}(t)$ will converge to $\bar{x}(t)$
exponentially fast. In view of the definitions of $\widehat{x}$ and $\bar{x}$, it is obvious that each $x_i$ will converge to $x$
 exponentially fast. So establishing convergence boils down to establishing the aforementioned properties of the $\Omega_j$ and $\Theta_j$
  sequences. For this a suitably defined matrix norm is needed. Such a norm, termed a mixed-matrix norm,''
  has been defined before
  \cite{ShaoshuaiTAC2015} and is described below.

Let  $|\cdot |_{\infty}$ denote   the standard induced infinity norm
  and write $\R^{mn\times mn}$ for  the vector space of
all $m\times m$  block matrices $M = \matt{M_{ij}}$ whose $ij$th
entry     is a matrix $M_{ij}\in\R^{n\times n}$.
 As in \cite{ShaoshuaiTAC2015} we define the {\em mixed matrix norm } of $M\in\R^{mn\times mn}$, written $||M||$, to be
\eq{||M|| = |\langle M\rangle |_{\infty}\label{mmn}}
 where $\langle M\rangle $ is the  matrix in $\R^{m\times m}$  whose $ij$th entry is $|M_{ij}|_2$.
It is very easy to verify that $||\cdot ||$ is in fact a norm. It is
even sub-multiplicative \cite{ShaoshuaiTAC2015}.

Corollary 1 of \cite{ShaoshuaiTAC2015} and its proof imply the following.
\begin{proposition}\label{prop:contraction}
Let $P_1,P_2,\ldots, P_m$ be any set of $n\times n$ orthogonal
matrices for which $\cap_{i\in\mathbf{m}}\ker P_i = 0.$ Let
 $\mathbb{N}_1,\; \mathbb{N}_2,\; \cdots,\;
\mathbb{N}_{(m-1)^2}$ be any sequence of self-arced, strongly
connected, directed graphs on $m$ vertices;  for $i\in\mathbf{m}$,
write $F_i$ for the flocking matrix $F_i = D^{-1}_iA_i'$ where $D_i$
is the diagonal matrix of in-degrees of vertices of  $\mathbb{N}_i$
and $A_i$ is the adjacency matrix of $\mathbb{N}_i$. Let
$\mathcal{C}$ denote the compact set of products of form $P_{j_1},
P_{j_2},\cdots, P_{j_{(m-1)^2}}$ where each of the $P_i$, $i\in
\mathbf{m}$, occurs in the product at least once. Then,
\eq{\label{eq:Mnorm} ||P(F_{(m-1)^2}\otimes I)P(F_{(m-1)^2-1}\otimes
I)\cdots P (F_1\otimes I )P|| \leq \gamma} where \eq{\gamma =
1-\frac{(m-1)(1-\rho)}{m^{(m-1)^2}}\label{gamma}} and $\rho=
\max\limits_{\mathcal{C}} |P_{j_1}P_{j_2}\cdots P_{j_{(m-1)^2}}|_2$.
Moreover,  $\rho <1$ and $\gamma <1$.
\end{proposition}

With Proposition~\ref{prop:contraction}, the following property of
the $\Omega_j$ sequence can be derived.

\begin{lemma}\label{lemma:1}
Let $\zeta$ be the largest eigenvalue of matrix $\frac{1}{2}(A+A')$.
Suppose that $q> (m-1)^2+1$ and that $\mathbb{N}(t)$ is
 a self-arced, strongly connected  neighbor graph for all $t\geq 0$. Then $\|\Omega_j\|\leq e^{\zeta T}\gamma^r$ where $r$ is the
  unique integer quotient of $q$ divided by $(m-1)^2+1$.
\end{lemma}

 \pf \textbf{of Lemma~\ref{lemma:1}:}
Since $\mathbb{N}(t)$ is strongly connected for all time, within
each time interval $[t_j-\tau,\; t_j)$, the graphs of the sequence
$\mathbb{N}(\tau_1),\; \mathbb{N}(\tau_2),\; \ldots,\;
\mathbb{N}(\tau_q)$ are all strongly connected. Also the graphs of
the sequence are all self-arced. By Proposition
\ref{prop:contraction} and sub-multiplicativity of the mixed-matrix
norm,
\begin{eqnarray*}
&&\|P(F_j((m-1)^2+c+1)\otimes I)\\ &&\hspace{0.2cm}
P(F_j((m+1)^2+c)\otimes I)\cdots P(F_i(c)\otimes I)\|\leq \gamma
\end{eqnarray*}
 for any positive integer $c$.
Thus we have
\begin{eqnarray} \nonumber &&
 \|P(F_j(q)\otimes I)P(F_j(q-1)\otimes I)\\ && \hspace{2cm}\cdots P(F_j(1)\otimes I )\| \leq \gamma^{r}
 \label{eq: Rinequality}
\end{eqnarray}
By \cite{HUexpnorm}, we know
 $
 |e^{At}|_2\leq e^{\zeta t}$ which means
 \begin{equation}
 \label{eq:eATinequality}
 \|e^{\bar{A}t}\|=|e^{At}|_2\leq e^{\zeta t}
 \end{equation}

By combining \rep{eq: Rinequality} and \rep{eq:eATinequality}, we
get
\begin{eqnarray*}
 \|\Omega_i\|  & \leq &
e^{\zeta\tau} \gamma^{r} e^{\zeta (T-\tau)} \\
& \leq & e^{\zeta T} \gamma^{r}
\end{eqnarray*} which completes the proof.
$\qed$ \\
 Now, with Lemma \ref{lemma:1}, convergence from $\hat{x}$
to $\bar{x}$ can be derived.

\pf \textbf{of Theorem~\ref{thm:1}:} As a first step, we find the
constraint for $q$ such that $\|\Omega_j\|<1$. By Lemma
\ref{lemma:1}, $\|\Omega_j\|\leq e^{\zeta T} \gamma^{r} \dfb \beta $
where $\zeta$ is the largest eigenvalue of $\frac{1}{2}(A+A')$. It
is sufficient to ensure $\beta<1$. Thus we get that if $q$ holds for
\rep{q}, $\beta<1$ holds.
If all the eigenvalue of $A$ is negative,  the right hand
side of the above inequality is less than zero and  it is trivial to
set $q=0 $.

Second, since $\mu(t)$ converges to zero at a pre-assigned rate, let
$\|\mu(t) \|\leq \bar \|\mu(0)\|g_1 e^{-wt}$ where $w$ can be any
positive numbers, and $\bar g_1$ is a positive constant. That is
equivalent to $\| \mu(t_j-\tau)\| \leq
\|\mu(0)\|g_1\alpha^{\frac{\tau}{T}+j}$, where $\alpha =e^{-wT}$ and
$g_1$ is positive.

Choose $w$ such that $\alpha<\beta$. From \rep{eq:errordynamic}, we
get
\begin{eqnarray*}
\hat{x}(t_j)-\bar{x}(t_j) & = &
\Omega_j\Omega_{j-1}\cdots\Omega_1(\hat{x}(0)-\bar{x}(0))
\\ &\ & +\sum_{s=1}^j\Omega_j\Omega_{j-1}\cdots\Omega_{s+1}\Theta_s\mu(t_s-\tau)
\end{eqnarray*}
 Let  $g_2\dfb\max _s \|\Theta_s\|$.
Then
\begin{eqnarray*}
\|\hat{x}(t_j)-\bar{x}(t_j)\| & \leq &
\|\hat{x}(0)-\bar{x}(0)\|\|\Omega_j\Omega_{j-1}\cdots\Omega_1\|
\\ &\ & \hspace{-0.8in} +\|\mu(0)\|g_1g_2
\sum_{s=1}^j\|\Omega_j\Omega_{j-1}\cdots\Omega_{s+1}\|\alpha^{\frac{T}{\tau}+s}
\\&\ & \hspace{-1in}  \leq   \|\hat{x}(0)-\bar{x}(0)\| \beta^j +\|\mu(0)\|g_1g_2\sum_{s=1}^j\beta^{j-s}\alpha^{\frac{T}{\tau}+s}\\
&\ & \hspace{-1in} = \left(\|\hat{x}(0)-\bar{x}(0)\|
+\|\mu(0)\|\frac{g_1g_2}{\beta-\alpha}\right)\alpha^{\frac{T+\tau}{\tau}}\beta^j
\\&\ & \hspace{-0.8in}  -\|\mu(0)\|\frac{g_1g_2}{\beta-\alpha}\alpha^{\frac{T+\tau}{\tau}} \alpha^j
\\ &\ & \hspace{-1in} \leq   \left(\|\hat{x}(0)-\bar{x}(0)\| +\|\mu(0)\|\frac{g_1g_2}{\beta-\alpha}\alpha^{\frac{T+\tau}{\tau}}\right)\beta^j
\end{eqnarray*}
Since $\beta <1$, all $x_i(t_j)$ will converge to the real state
$x(t_j)$ exponentially fast whose rate is bounded by $\beta$. Thus
we can get the result that
\[|x_i(t)-x(t)|_2\leq e^{-\lambda t}(\sigma_x+\sigma_\mu), \;\; t\geq 0,\;\;i\in\mathbf{m}\]
where
\[\sigma_x=\|\hat{x}(0)-\bar{x}(0) \|,\;\; \sigma_\mu=\|\mu(0)\| \frac{g_1g_2}{\beta-\alpha}\alpha^{\frac{T+\tau}{\tau}}\]
and $\lambda=-\frac{1}{T}\ln \beta $ which is equivalent to
\rep{lambda}. Note $\|\mu(0)\|=\max \limits_{i\in \mathbf{m}}
|w_i(0)-L_ix(0)|_2$, and \[g\dfb
\frac{g_1g_2}{\beta-\alpha}\alpha^{\frac{T+\tau}{\tau}}\] is a
positive constant which completes the proof. $\qed$

\section{Concluding Remarks}

One of the nice properties of the hybrid observer discussed in this paper is that it is {\em resilient}.
By this we mean that under appropriate conditions it will  be able to continue to provide asymptotically correct estimates of $x$,
 even if one of the agents leave the network. To give an example of this, suppose
  that  $\mathbb{N}(t)$ is a constant, strongly connected, self-arced  four vertex  graph $\mathbb{N}$
  with arcs in both directions between each pair of
   vertices except for the  vertex pair $(1,4)$. Then $\mathbb{N}$ will remain  self-arced and strongly connected
after the removal of any one vertex  $v$ and all arcs
 either leaving or arriving at $v$. Suppose the system whose state is to be estimated  is a jointly observable
   four channel system
which remains jointly observable after any one measurement $y_i$ is removed.
    Suppose that agent $k$ leaves the network.  It is clear that
    any   properly designed   hybrid observer  for this system   will be  able
     generate asymptotically correct estimates of $x$  not only for  all four agents but also for the
      three  which remain  after agent $k$ leaves.
 Further research is needed to more fully understand resilient observers.

 Generally one would like to choose $T$ ``small'' to avoid unnecessarily large error overshooting between event times.
 Meanwhile, since $r$ is the unique integer quotient of $q$ divided by $(m-1)^2 +1$, it is obvious  from \rep{lambda}
that the
 larger the number of iterations $q$ on
  each interval $[t_j-\tau, t_j)$ the faster the convergence.
  Two considerations limit the value of $q$ - how fast
  the parameter estimators can compute  and how quickly information can be transmitted across the network. We doubt the former
  consideration will prove very important in most applications, since digital processers can be quite fast and the
   computations required
   are not so taxing.  On the other hand, transmission delays will almost certainly limit the choice of  $q$.
   A model which explicitly takes such delays into account will be presented in another paper.

   A second practical issue which this paper does not address is the question of synchronization.  Like just about all
   published papers on distributed estimation, this paper implicitly assumes that all agents share a common clock.
    This is undoubtedly an unrealistic assumption for many applications. The tools to study this type of observer
     in an asynchronous setting
     already exist \cite{async.liu} and we anticipate applying them  in the future.

     A third practical issue is that the development in this paper does not take into account measurement noise.
     On the other hand, the observer provides exponential convergence and this suggests that if noisy measurements are considered, the system
      the observer's performance will degrade gracefully with increasing noise levels. Of course one would like an ``optimal''
      estimator for such situations in the spirit of a Kalman filter. Just how to formulate and solve such a problem
        is a significant issue
      for further research.

\bibliographystyle{unsrt}
\bibliography{my,steve,references}

\end{document}